\documentclass[a4paper,11pt,onecolumn]{article}

\usepackage[dvipdfmx]{graphicx}
\usepackage[utf8]{inputenc}
\usepackage[T1]{fontenc}
\usepackage{amsmath, amssymb, amsxtra}
\usepackage{nccmath, icomma, xspace, bbold}
\usepackage[mathic]{mathtools}
\usepackage[version=3]{mhchem}
\usepackage{siunitx}
\DeclareSIUnit\krec{\ensuremath{\textit{k}_\text{rec}}}
\DeclareSIUnit\Erec{\ensuremath{\textit{E}_\text{rec}}}
\usepackage[short]{datetime}
\usepackage[normalem]{ulem}
\usepackage{color}

\usepackage[%
	left = 1.5cm,%
	right = 1.5cm,%
	top = 1cm,%
	bottom = 1.5cm,%
	footskip = 0.5cm,%
	nohead]
	{geometry}
\usepackage{nccparskip}
\SetParskip{4pt}
\usepackage{setspace}
\usepackage{enumitem}

\usepackage[british]{babel}
\usepackage{csquotes}

\usepackage{ctable}
\usepackage{caption}%
\DeclareCaptionLabelSeparator{vertline}{~|~}%
\usepackage{authblk}

\usepackage{xifthen}
\newcommand{\ket}[2][]{ %
	\ifthenelse{\isempty{#1}}%
	{\ensuremath{\xspace\left\vert #2 \right\rangle\xspace}}%
	{\ensuremath{\xspace\left\vert #2 \right\rangle_{\! #1}\xspace}}
	}
\newcommand{\bra}[2][]{ %
	\ifthenelse{\isempty{#1}}%
	{\ensuremath{\xspace\left\langle #2 \right\vert\xspace}}
	{\ensuremath{\xspace\prescript{}{#1}{\!\left\langle #2 \right\vert\xspace}}}
	}
\newcommand{\braket}[3][]{ %
	\ifthenelse{\isempty{#1}}%
	{\ensuremath{\xspace\left\langle #2 \left\vert\right. #3 \right\rangle\xspace}}
	{\ensuremath{\xspace\left\langle #2 \left\vert\right. #3 \right\rangle_{\! #1}\xspace}}
	}
\newcommand{\avg}[2][]{ %
	\ifthenelse{\isempty{#1}}%
	{\ensuremath{\xspace\langle #2 \rangle\xspace}}%
	{\ensuremath{\xspace\langle #2 \rangle_{\! #1}\xspace}}
	}

\newcommand{\krec}{\ensuremath{k_\text{rec}}\xspace}
\newcommand{\Erec}{\ensuremath{E_\text{rec}}\xspace}

\begin{document}
\title{
Frequency-domain Hong-Ou-Mandel interference
}
\author[1]{Toshiki~Kobayashi}
\author[1]{Rikizo~Ikuta}
\author[1]{Shuto~Yasui}
\author[2]{Shigehito~Miki} 
\author[2]{ Taro~Yamashita}
\author[2]{Hirotaka~Terai}
\author[1]{Takashi~Yamamoto}
\author[3]{Masato~Koashi}
\author[1]{Nobuyuki~Imoto}
\affil[1]{Graduate School of Engineering Science, Osaka University,
Toyonaka, Osaka 560-8531, Japan}
\affil[2]{Advanced ICT Research Institute, 
National Institute of Information and Communications Technology (NICT),
Kobe 651-2492, Japan}
\affil[3] {Photon Science Center, Graduate School of Engineering,
The University of Tokyo, Bunkyo-ku, Tokyo 113-8656, Japan}
\date{}

\maketitle

Hong-Ou-Mandel (HOM) interference\cite{Ou1987} unveils a distinct behavior
of identical particles which cannot be distinguished from each other.
Especially for bosons,
two separated identical particles passing through a beamsplitter always go together
into one of the output ports, but that is not the case
with other particles including fermions or classical ones.
So far many elemental properties of quantum physics and information\cite{Pan2012}
have been discovered through the concatenated HOM effects, 
which has been demonstrated in photons
\cite{Ou1987, Santori2002, Patel2010, Beugnon2006, Maunz2007, Sipahigil2012, Sipahigil2014}
and recently in plasmons\cite{DiMartino2014, Fakonas2014}, atoms\cite{Lopes2015} and phonons\cite{Toyoda2015}. 
However, all demonstrations in optical region 
employed two particles in different spatial modes. 
Here we first report the HOM interference between two photons 
in a single spatial mode with different frequencies (energies)
by using a partial frequency conversion.
The demonstrated frequency-domain interferometer allows us
to replace spatial optical paths by optical frequency multiplexing,
which opens up a distinct architecture of the quantum interferometry.

In the past three decades since the HOM interference has been proposed and demonstrated with two photons from spontaneous parametric down-conversion (SPDC) process \cite{Ou1987}, a huge varieties of experiments based on the HOM interference revealed fundamental properties in quantum physics, especially in quantum optics\cite{Pan2012}, and  its applications are widely spreading over quantum information processing, such as quantum computation\cite{Kok2007, Barz2012, Spagnolo2014, Carolan2015}, quantum key distribution\cite{Tang2014, Guan2015}, quantum repeater\cite{Sangouard2011, Hofmann2012, Bao2012} and quantum-optical coherence tomography\cite{Teich2012}. HOM interference has been observed with photons generated not only from  nonlinear optical phenomenon but also from quantum dots\cite{Santori2002, Patel2010}, trapped neutral atoms\cite{Beugnon2006},
trapped ions\cite{Maunz2007}, NV centers\cite{Sipahigil2012} and SiV centers\cite{Sipahigil2014} in diamond. 
Furthermore not only photons but also other bosonic particles, e.g., surface plasmons\cite{DiMartino2014, Fakonas2014},  Helium 4 atoms\cite{Lopes2015} and phonons\cite{Toyoda2015} show the HOM interference.
In spite of such demonstrations using various kinds of physical systems, 
to the best of our knowledge,
all of them essentially used the spatial or polarization degree of freedom for the HOM interference, 
including the use of polarization modes of photons that are easily converted to and from spatial modes.
The demonstrations use the beamsplitter (BS) which mixes the two particles in different spatial/polarization modes.

\begin{figure*}[t]
 \begin{center}
  \includegraphics[bb=0 0 485 395, width=0.9\linewidth]{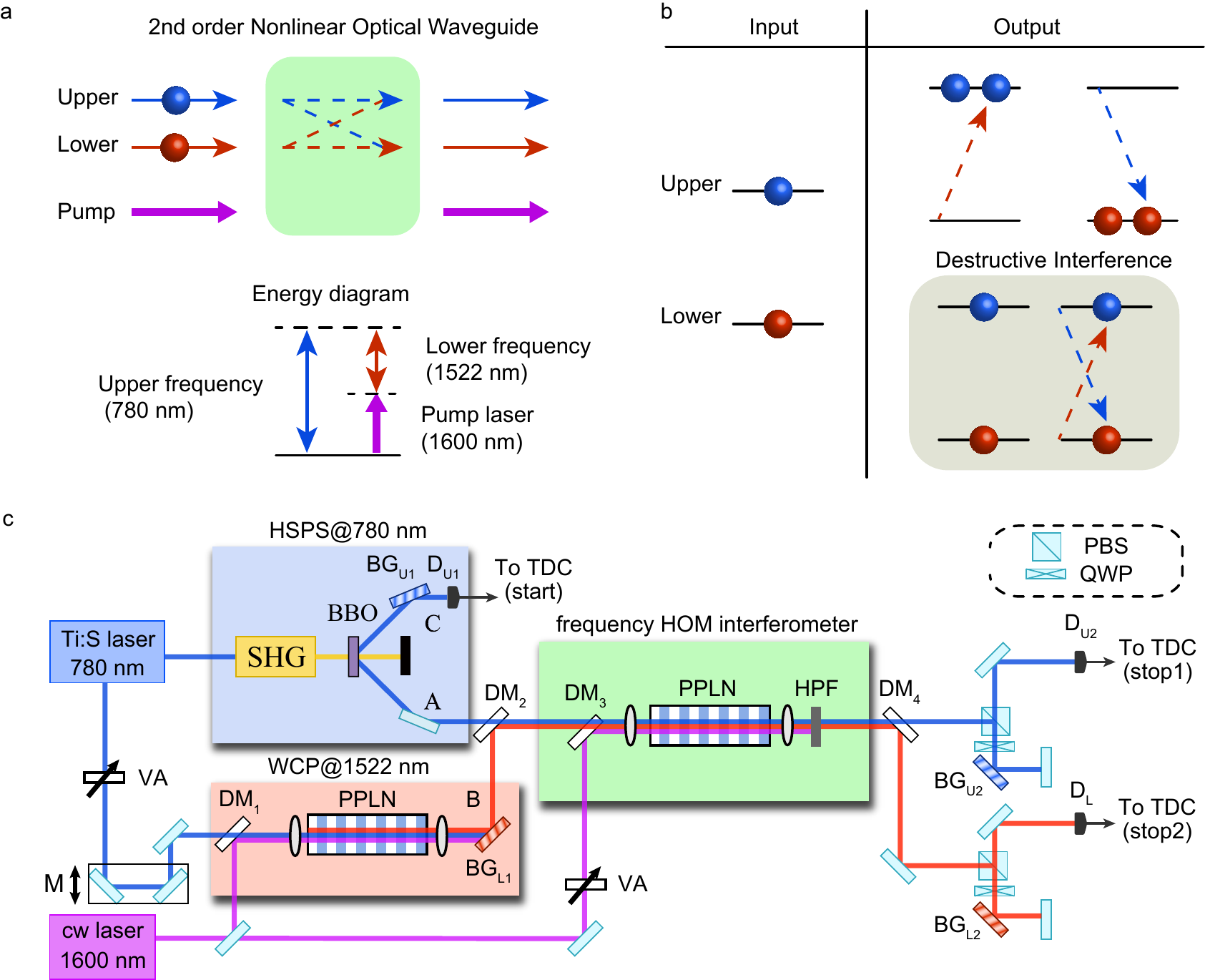}
 \end{center}
 \caption{{  Frequency-domain HOM interferometer.}
 { a},  Frequency converter based on second-order nonlinear optical effect. 
 It partially converts the wavelengths of the photons in a single spatial mode from/to 780~nm to/from 1522~nm
 via sum/difference frequency generation.
 { b}, Principle of the frequency-domain HOM effect. 
  When a single photon in upper mode and another photon in the lower mode are mixed by the frequency converter,
  the single photon occupation events in the output disappear due to the  destructive interference. 
 { c}, The experimental setup of the frequency-domain HOM interference.
  In the experiment,  the heralded single photon source~(HSPS) at 780~nm and
   the weak coherent pulse~(WCP) at 1522~nm are prepared
    to serve as two input photons to the frequency HOM interferometer.
 The two photons are combined by $\mathrm{DM_2}$ 
 to a single spatial mode and then go into the PPLN waveguide as the frequency-domain BS. 
 The output light pulses are separated into two spatial modes by $\mathrm{DM_4}$  for the photon detection of the two frequency modes. 
 }
 \label{fig:setup}
\end{figure*}

In this letter, we report the first observation of the HOM interference between two photons
with different frequencies in optical region.
In contrast to the spatial interferometer, 
the frequency-domain HOM interferometer is implemented in 
a single spatial mode with a nonlinear optical frequency conversion\cite{Tanzilli2005, Ikuta2011, Ikuta2013}.
In the experiment, we input a 780 nm photon and a 1522 nm photon to the frequency converter 
that partially converts the wavelengths of the photons between 780~nm and 1522~nm
as shown in Fig.~\ref{fig:setup}a. 
We measured coincidence counts between the output photons at 780~nm and those at 1522~nm 
from the frequency converter.
The observed HOM interference between the two photons
 in a single spatial mode at different frequencies 
 clearly indicates the nonclassical property. 

The frequency-domain HOM interference between the photons with different frequencies
in this paper is performed by using a partial frequency converter based on the second-order nonlinear optical effect \cite{Ikuta2011, Kumar1990}. 
Suppose that an upper angular frequency $\omega_s$, a lower angular frequency $\omega_i$   
and an angular frequency $\omega_p$ of the pump light satisfy $\omega_i=\omega_s-\omega_p$.
When the pump light is sufficiently strong, the effective Hamiltonian of the nonlinear optical process is described by
\begin{equation}
  \label{eq:h}
  \hat{H}=i\hbar \left(g^* \hat{a}_i^\dagger \hat{a}_s-g\hat{a}_i \hat{a}_s^\dagger \right),
\end{equation}
where $\hat{a}_s$ and $\hat{a}_i$ are annihilation operators of the upper and the lower frequency modes, respectively.
$g = |g|e^{i\phi}$ is proportional to the complex amplitude of the pump light, 
where $\phi$ represents the phase of the pump light.
By using Eq.~(\ref{eq:h}), annihilation operators $\hat{a}_{s,\mathrm{out}}$ and $\hat{a}_{i,\mathrm{out}}$
of the signal and the idler modes coming out from the nonlinear optical medium are described by
\begin{equation}
  \label{eq:signal}
  \hat{a}_{s,\mathrm{out}}= \cos(|g|\tau)\hat{a}_s-e^{i\phi}\sin(|g|\tau)\hat{a}_i
\end{equation}
and
\begin{equation}
  \label{eq:idler}
  \hat{a}_{i,\mathrm{out}}= e^{-i\phi} \sin(|g|\tau)\hat{a}_s+\cos(|g|\tau)\hat{a}_i,
\end{equation}
where $\tau$ is the traveling time of the light pulses through the nonlinear optical medium.
The probability of a photon staying in the same input frequency mode is given by $\cos^2(|g|\tau)$ and the transition probability of the photon from an input frequency mode to the other frequency mode is $\sin^2(|g|\tau)$. 
The process of the frequency conversion described in Eqs.~(\ref{eq:signal}) and (\ref{eq:idler}) can be seen as a BS with two different frequency input (output) modes\cite{Giorgi2003, Raymer2010, Ikuta2013:3}. 
Borrowing the terminology of the spatial BS, we may regard $\cos^2(|g|\tau)$ and $\sin^2(|g|\tau)$ as  
the transmittance and the reflectance, respectively. 

In the frequency converter, the transition probability can be adjusted by changing the pump power.
When we choose the pump power such that $\cos^2(|g|\tau)=\sin^2(|g|\tau)=1/2$, the frequency converter works as a half BS acting on the two frequency modes. 
In such a situation, when a single photon in upper mode and another single photon in lower mode are injected to the frequency converter simultaneously, the two photons never come out from the different frequency modes but 
always come out in the same frequency modes.
This phenomenon is a precise analog of the HOM interference in the frequency degree of freedom. 
One may wonder why unlike the conventional HOM experiments, two distinct bosons, which are distinguished by their frequencies, show the HOM interference. As shown in Fig.~\ref{fig:setup}b, however, the crux of the HOM interference is the destructive interference between the output events. 
Therefore, what is really required in the process is the indistinguishability after the frequency conversion, which can be fulfilled by a suitable coherent property of the converter.

\begin{figure}[t]
 \begin{center}
  \includegraphics[bb=0 0 565 187, width=0.9\linewidth]{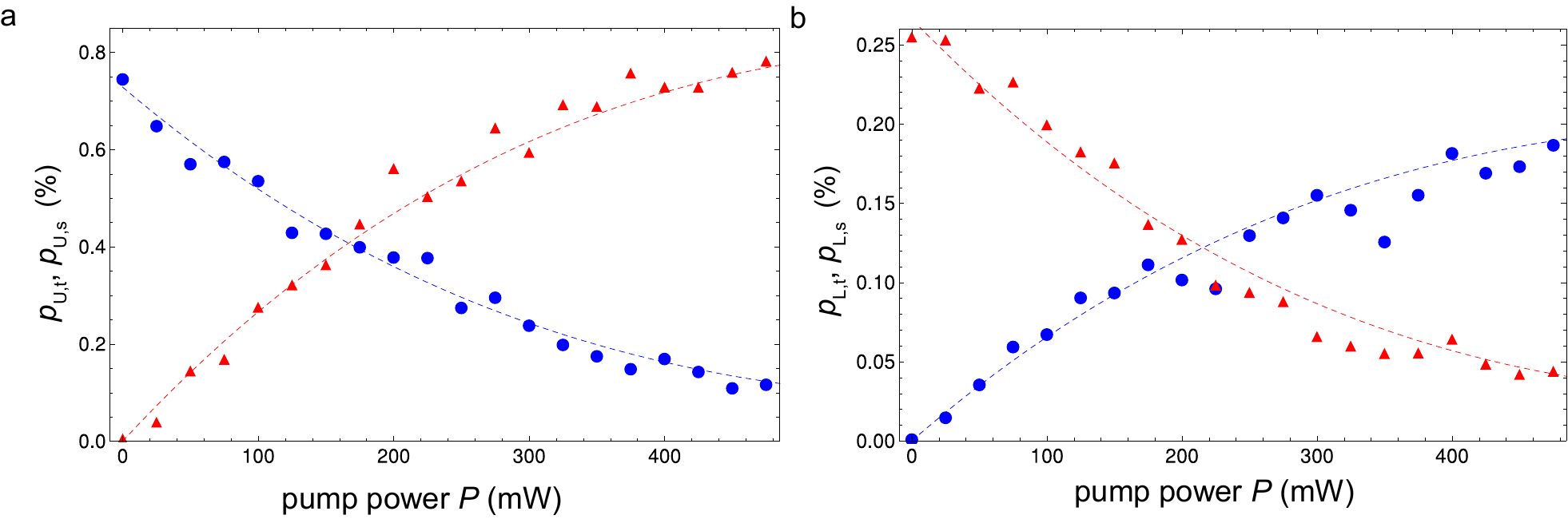}
 \end{center}
 \caption{{ The count rate vs. the pump power}.
 { a},  The count rate of the transition/staying events per pulse 
 	$p_{\mathrm{U,t/s}}$ (circle/triangle) 
 	for the heralded single photon at 780~nm 
         when the heralding signal is detected at $\mathrm{D}_{\mathrm{V1}}$. 
   { b}, The count rate of the transition/staying events per pulse
   	$p_{\mathrm{L,t/s}}$ (triangle/circle)
   	for the coherent light pulse at 1522~nm. 
   	The dashed curves are obtained from our theoretical model with the observed values of 
   	$p_{\mathrm{U,t/s}}$ and $p_{\mathrm{L,t/s}}$ (see Supplementary material).
 }
 \label{fig:pre}
\end{figure}
\begin{figure}[t]
 \begin{center}
  \includegraphics[bb=0 5 287 187, width=0.6\linewidth]{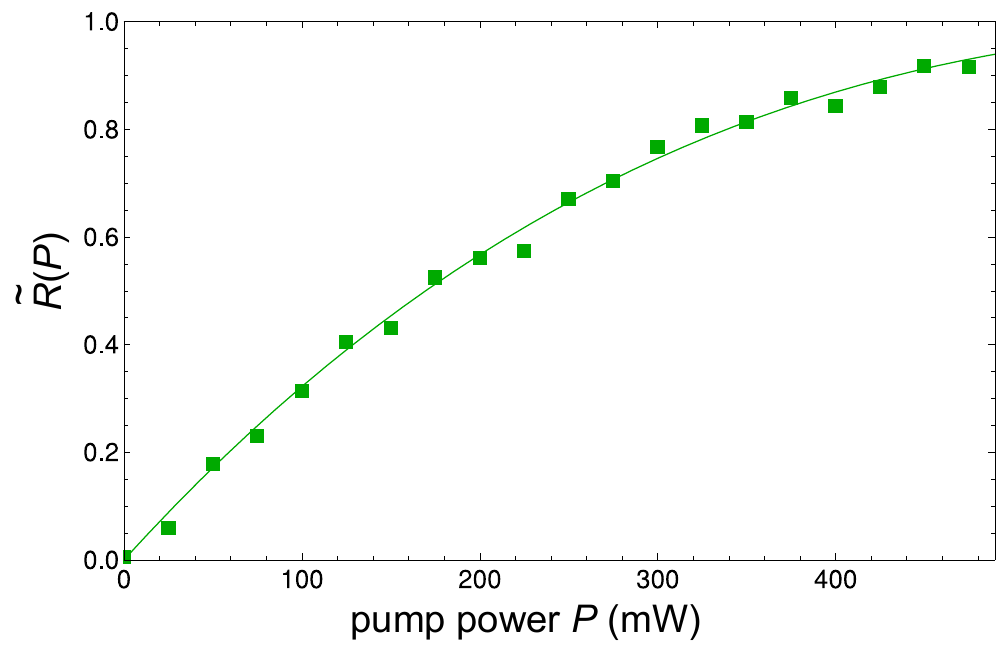}
 \end{center}
 \caption{{ The peak value of the internal transition probability.}
   The curve is obtained by the best fit to $\widetilde{R}(P)$ with $A \sin ^2(\sqrt{\eta P})$, where
     $A \approx 0.99$ and $\eta \approx0.0036$~/mW.
 }
 \label{fig:conv}
\end{figure}

The experimental setup for the frequency-domain HOM interference by using
the partial frequency converter is shown in Fig.~\ref{fig:setup}c.
We prepare a vertically(V) polarized heralded single photon at 780~nm in mode A
 and a V polarized weak coherent light at 1522~nm in mode B with an average photon number of 0.1 (see Method).
The two light pulses are 
combined by a dichroic mirror~($\mathrm{DM_2}$) 
and then focused on a  type-0 quasi-phase-matched periodically-poled LiNbO${}_3$~(PPLN) waveguide\cite{Ikuta2011} for the frequency conversion.
The time difference between the two light pulses is adjusted by mirrors~(M) on a motorized stage.
The V polarized cw pump laser at 1600 nm is combined with the two input light pulses by $\mathrm{DM_3}$ and focused on the PPLN waveguide. 
The length of the PPLN crystal is 20 mm and the acceptable bandwidth is calculated to be
$\Delta_{\mathrm{WG}}\equiv 140$~GHz which corresponds to 0.28~nm for 780-nm light and 1.1~nm for 1522-nm light. 
The pump power is adjusted by a variable attenuator~(VA) 
and determines the transition probability of the frequency converter.

After the frequency converter, the light pulses at 780 nm and 1522 nm are separated by $\mathrm{DM_4}$ and
Bragg gratings ($\mathrm{BG_{U2}}$ and $\mathrm{BG_{L2}}$ with  bandwidths of $\Delta_{\mathrm{U}} \equiv 99$~GHz and $\Delta_{\mathrm{L}} \equiv 130$~GHz, respectively). 
They are then measured by an avalanche photodiode 
with the quantum efficiency of about 60\% for 780-nm photons ($\mathrm{D_{U2}}$) and
 by a superconducting single-photon detector~(SSPD)\cite{Miki2013}  with the quantum efficiency of about 60\% for the 1522-nm photons ($\mathrm{D_{L}}$), respectively.
In order to observe the HOM interference, we collect the threefold coincidence events among
the three detectors $\mathrm{D_{U1},D_{U2}}$ and $\mathrm{D_{L}}$.
Note that SSPD with the quantum efficiency of about 10\%
 for 780-nm photons ($\mathrm{D_{U1}}$) is used for heralding the 780-nm input photon in mode A.

Before we demonstrate the frequency-domain HOM interference, 
we first measured the dependencies of
the count rates of the transition/staying events $p_{\mathrm{U, t/s}}$ and $p_{\mathrm{L, t/s}}$
on the pump power for each of the upper and the lower input photons, respectively.
The experimental result is shown in Figs.~\ref{fig:pre}a and \ref{fig:pre}b.
From the experimental result, 
we estimate the internal transition probability $R$ of the frequency converter, 
which depends on the pump power $P$ and the frequency of the input light,
by constructing a theoretical model as follows. 

\begin{figure*}[t]
 \begin{center}
  \includegraphics[bb=0 0 534 184, width=0.9\linewidth]{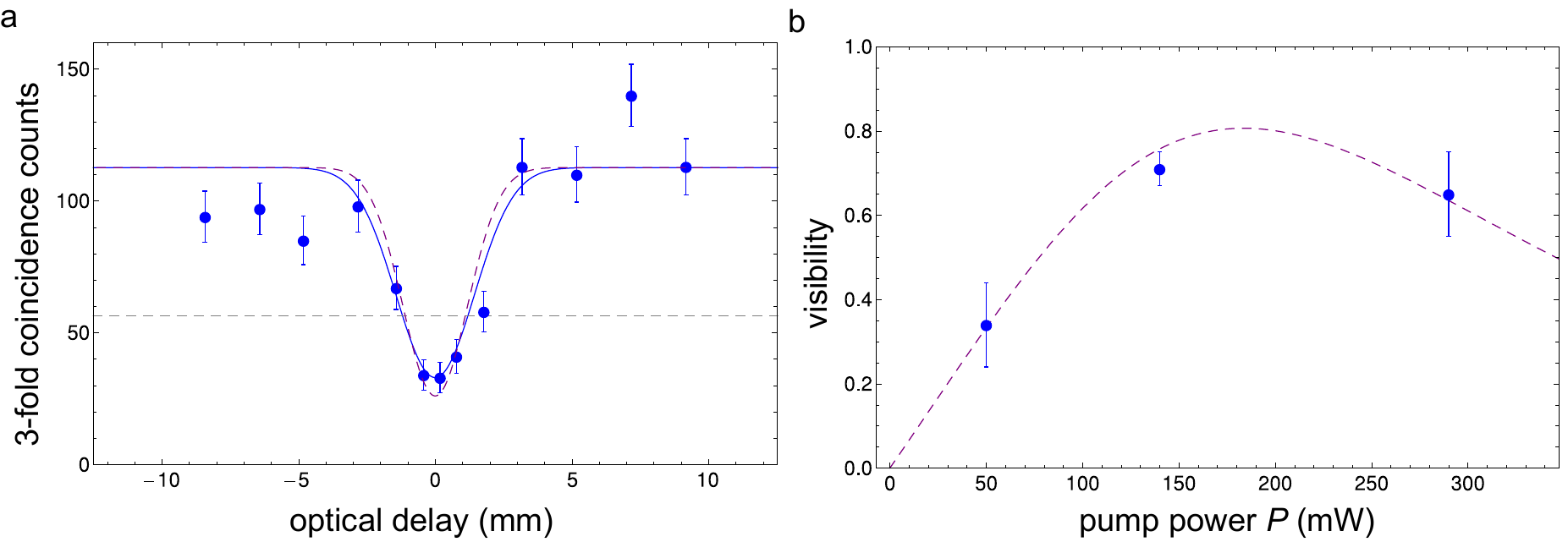}
 \end{center}
 \caption{{ Observed frequency-domain HOM interference. } 
  { a},  The observed HOM dip at 140-mW pump power. 
  The circles represent the experimental threefold coincidence counts.
  The solid curve is the Gaussian fit to the experimental counts.
  The dashed curve is obtained from our theoretical model with the experimental parameters.
  The dashed horizontal line describes 
  the half values of the maximum of the fitting result. 
   { b},
  The pump power dependence of the visibility.
  The circles are obtained from the experimental result.
    The dashed curve is obtained from our theoretical model with the experimental parameters.
    }
 \label{fig:dip}
\end{figure*}
We assume that the internal transition probability $R=R(P, \omega)$ is a 
Gaussian with the bandwidth of $\Delta_{\mathrm{WG}}$ around
the center of 780~nm/1522~nm for upper/lower input light,
at which the peak value is given by $\widetilde{R}(P)$.
We also assume that 
the optical circuit for the frequency-domain HOM after preparing the single photon and the coherent light pulse
is decomposed to a lossless frequency converter with the transition probability $R(P, \omega)$ and the staying probability $T(P, \omega)(=1-R(P, \omega))$, two lossy media inducing the loss in each of the upper and the lower input modes, and two spectral filters acting on two output modes.
The transmittances of the lossy media for the upper and the lower input light are denoted by $T_{\mathrm{in, U}}$ and $T_{\mathrm{in, L}}$, respectively, which describe the optical loss including the insertion loss to the frequency converter. 
The transmittances of the spectral filters for the upper and the lower output light are denoted by 
$T_{\mathrm{out, U}}(\omega)$ and $T_{\mathrm{out, L}}(\omega)$, respectively, which describe 
the optical loss including the BGs after the frequency converter
and the quantum efficiencies of the detectors.
We assume that the transmittance $T_{\mathrm{out, U/L}}(\omega)$ is Gaussian with
the peak value of $\widetilde{T}_{\mathrm{out, U/L}}$.
The bandwidths of $T_{\mathrm{out, U}}(\omega)$ and $T_{\mathrm{out, L}}(\omega)$ are calculated to be 70~GHz and 92~GHz, respectively, from the effect of using the $\mathrm{BG_{U2}}$ and $\mathrm{BG_{L2}}$ twice.
For the input light pulses, we assume that the spectral shapes of the heralded single photon and the coherent light pulse
 are Gaussian with bandwidths of 740~GHz and 93~GHz, respectively.
These are calculated by using the experimental parameters  $\Delta_{\mathrm{WG}}, \Delta_{\mathrm{U}}$, $\Delta_{\mathrm{L}}$ and the pulse width of the laser source $\Delta t \equiv 1.2$~ps.
The emission rate of the photon pair from the SPDC is so small that the multiple-pair events
are negligible. 
In addition, the heralded genuine single photon is assumed to be 
in a pure state because the narrow-band spectral filtering destroys the spectral correlation between
the photon pair from SPDC.
Under these assumptions, the initial state composed of the single photon and the coherent state
is regarded as a pure state described by $\ket{\phi}=\hat{a}^\dagger_{\mathrm{U}}\hat{D}_{\mathrm{L}}(\alpha)\ket{0}$, 
where $\hat{a}^\dagger_{\mathrm{U}}$ is a creation operator of the upper mode,
 $\hat{D}_{\mathrm{L}}(\alpha)$ is a displacement operator with complex number $\alpha$ of the lower mode
  and $\ket{0}$ is the vacuum state for both modes.
Based on the above theoretical model with the use of 
the observed values $p_{\mathrm{U,t}}(P), p_{\mathrm{U,s}}(P), p_{\mathrm{L,t}}(P)$ and $p_{\mathrm{L,s}}(P)$,
we calculated the four parameters $\widetilde{R}(P)$,  $T_{\mathrm{in, U}}\widetilde{T}_{\mathrm{out, U}}$,
 $T_{\mathrm{in, U}}\widetilde{T}_{\mathrm{out, L}}$ and $|\alpha|^2 T_{\mathrm{in, L}}/T_{\mathrm{in, U}}$
  for various values of $P$ (see Supplementary material). 
The result of $\widetilde{R}(P)$ is shown in Fig.~\ref{fig:conv}.

Next we demonstrated the frequency-domain HOM interference
by using the frequency converter.
We set a pump power to be 140~mW which results in the transition probability $\sim0.4$ of the frequency converter according to Fig.~\ref{fig:conv}.
The experimental result of the dependency of the threefold coincidence counts
on the optical delay is shown in Fig.~\ref{fig:dip}a. 
The observed visibility of $0.71\pm 0.04$ at the zero delay point
was obtained by the best fit to the experimental data with a Gaussian.
The full width at the half maximum was approximately $1.7$~mm which corresponds to
$\sim 6$~ps of a delay time.
The high visibility clearly shows the nonclassical HOM interference between
the two light pulses in a single spatial mode with different frequencies.
We also measured the visibilities at the pump power 50~mW and 290~mW, which corresponds to
the transition probabilities $\sim 0.2$ and $\sim 0.7$, respectively.
The experimental result is shown in Fig.~\ref{fig:dip}b.
The observed visibilities are $0.34\pm 0.10$ at 50~mW and $0.65 \pm 0.10$ at 290~mW.

In the following, we discuss the reasons for the degradation of the visibility.
In our theoretical model, 
we can calculate the visibility
by using the experimental parameters $\widetilde{R}(P_0)$,  $T_{\mathrm{in, U}}\widetilde{T}_{\mathrm{out, U}}$,
 $T_{\mathrm{in, U}}\widetilde{T}_{\mathrm{out, L}}$ and $|\alpha|^2 T_{\mathrm{in, L}}/T_{\mathrm{in, U}}$
 (see Supplementary material),
the results of which are the dashed curves shown in Figs.~\ref{fig:dip}a and ~\ref{fig:dip}b.
These are in good agreement with the experimental results.
Fig.~\ref{fig:dip}b indicates that the highest visibility of 0.81 will be obtained 
when the pump power is 190~mW which corresponds to the transition probability $\sim 0.5$.
In our theoretical model,
main reasons for the degradation of the visibility comes from the input light pulses;  
the effect of the multiphoton components in the coherent light pulse at 1522~nm
and the broad bandwidth of the heralded single photon at 780~nm.
If we replace the coherent light pulse by a single photon and set $T_\mathrm{in,L}$ to be equal to $T_\mathrm{in,U}$,
the visibility of the HOM interference is expected to be 0.95 at 190-mW pump power.
If we use the single photon at 780~nm with the same bandwidth as that of the coherent light pulse, 
the visibility will be 0.93 at 190-mW pump power. 
If we take both improvement for the input light pulses, 
the visibility will be 0.98 at 190-mW pump power.

In conclusion, we have demonstrated the frequency-domain HOM interference
between a heralded single photon at 780~nm and a weak laser light at 1522~nm in a single spatial mode 
by using the partial frequency converter based on the nonlinear optical effect. 
We also deduced that the performance of the demonstrated frequency domain HOM interferometer
is almost ideal from the fact that the estimated visibility in the case of the ideal input photons
is 0.98, which is close to unity.
So far the spatial HOM interferometer has been exploited in a wide variety of quantum phenomena
including a large scale quantum information processing.
We thus believe that the frequency domain HOM interferometer will open up a novel frequency domain quantum interferometry and give a novel tool for exploiting quantum phenomena and
a way of scaling up the quantum information processing with a large Hilbert space spanned by
widely spreading frequency modes. 

\section*{Methods}
{Preparation of the two input light pulses.}.
A light pulse from a mode-locked Ti:sapphire laser at 780~nm (pulse width: $\Delta t \equiv 1.2$~ps; 
repetition rate: 82~MHz) is used for the preparation. It is divided into two beams. 
One beam is used for preparing a heralded single photon at 780~nm.
The beam is frequency doubled (wavelength: 390~nm; power: 200~mW) by second-harmonic generation~(SHG),
and then pumps a type-I phase-matched 1.5-mm-thick $\beta$-barium borate~(BBO) crystal 
to generate a photon pair at 780~nm in modes A and C through the spontaneous parametric down conversion.
The photon in mode C is measured by a SSPD denoted by $\mathrm{D_{U1}}$,
 which prepares a heralded single photon in mode A.
The spectral filtering of the photon in mode C is performed by a Bragg grating ($\mathrm{BG_{U1}}$) with a bandwidth of $\Delta_{\mathrm{U}} \equiv 99$~GHz which corresponds to 0.2~nm for 780-nm light.

The other beam from Ti:S laser is used for preparing the weak coherent light pulse at 1522~nm.
The beam enters a difference frequency generation~(DFG) module.
In the DFG module, a V polarized cw pump laser at 1600~nm
is combined with the input light pulse at 780~nm by a dichroic mirror~($\mathrm{DM_1}$).
They are focused on a type-0 quasi-phase-matched PPLN waveguide\cite{Ikuta2011}.
The length of the PPLN crystal is 20 mm and the acceptable bandwidth is calculated to be
$\Delta_{\mathrm{WG}}= 140$~GHz.
After passing through the PPLN waveguide, the converted light at 1522~nm is 
extracted by $\mathrm{BG_{L1}}$ with a bandwidth of $\Delta_{\mathrm{L}}\equiv 130$~GHz which corresponds to 1~nm for 1522-nm light.
We adjust the average photon number of the coherent light pulse at 1522~nm
to be $\sim 0.1$ by a variable attenuator~(VA).

\section*{Acknowledgements}
This work was supported by JSPS Grant-in-Aid for JSPS Fellows 14J04677,
Scientific Research(A) 25247068, (B) 15H03704 and (B) 25286077.

\newpage
\setcounter{figure}{0}
\section*{Supplementary material}
\renewcommand{\thefigure}{S\arabic{figure}}
\begin{figure}[n]
 \begin{center}
  \includegraphics[bb=0 0 316 167, width=0.8\linewidth]{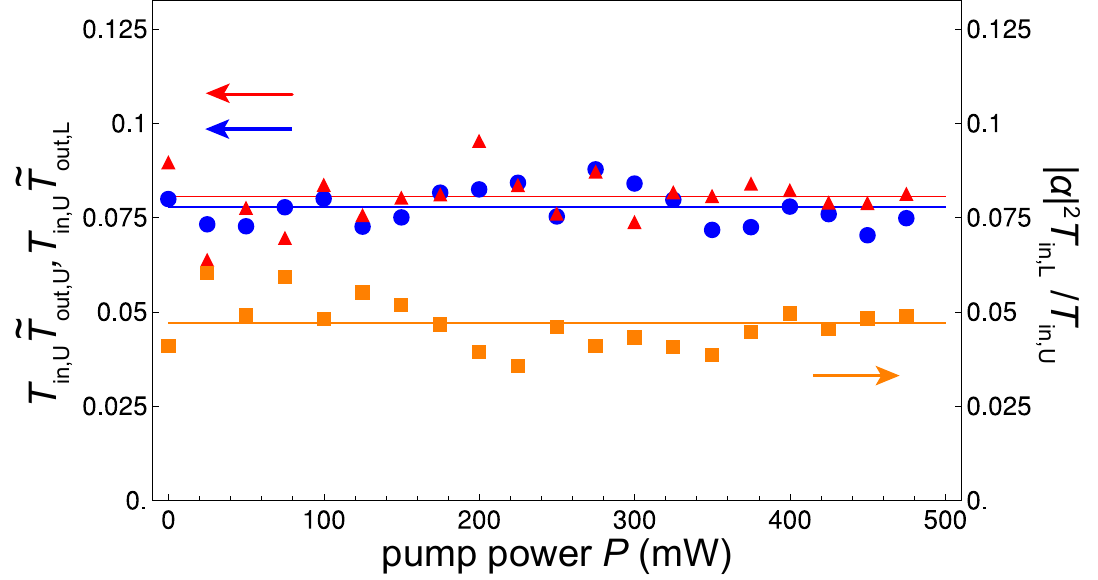}
 \end{center}
 \caption{{ The estimated values of the parameters except for the internal transition probability vs. the pump power.}
 The circles, triangles and squares indicate the values of $T_{\mathrm{in, U}}\widetilde{T}_{\mathrm{out, U}}$,
 $T_{\mathrm{in, U}}\widetilde{T}_{\mathrm{out, L}}$ and $|\alpha|^2 T_{\mathrm{in, L}}/T_{\mathrm{in, U}}$,
 respectively.
 The horizontal lines represent the average values of these three parameters.
 }
 \label{fig:precalc}
\end{figure}

\begin{figure}[n]
 \begin{center}
  \includegraphics[bb=0 0 276 172, width=0.7\linewidth]{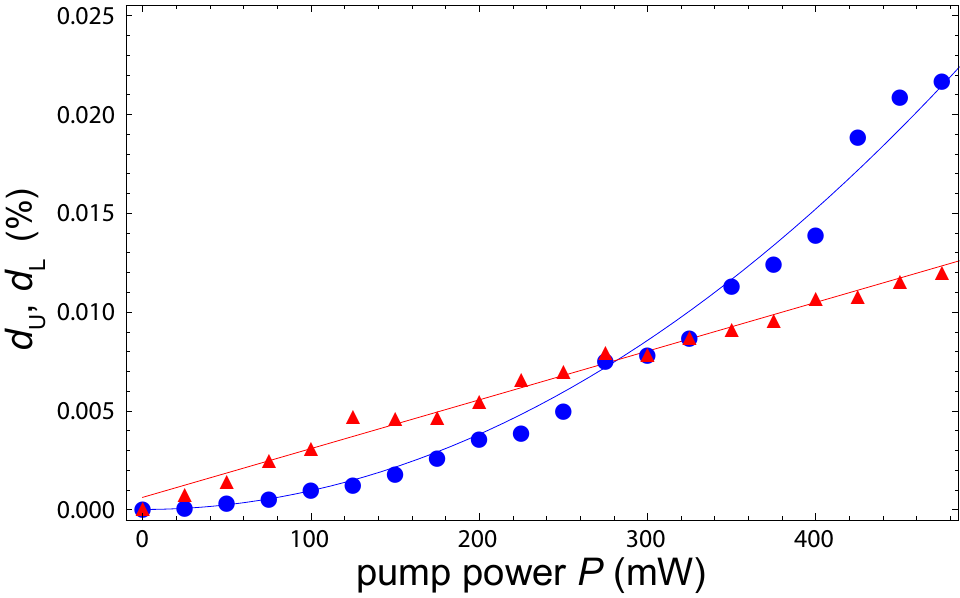}
 \end{center}
 \caption{{ The observed detection probability of the background noise.}
 The circles/triangles indicate the observed detection probabilities of the background noise  $d_{\mathrm{U/L}}(P)$.
 The curve for $d_{\mathrm{U}}(P)$ is obtained by the best fit to the observed values with 
 $A P^2 + B P + C$, where $A \approx 9.5 \times 10^{-8}/\mathrm{mW}^2, B\approx 0.0 /\mathrm{mW}$ and $C\approx 0.0$.
 The curve for $d_{\mathrm{L}}(P)$ is obtained by the best fit to the observed values with 
 $D P + E$, where $D \approx 2.5 \times 10^{-5}/\mathrm{mW}$ and $E\approx 6.1\times 10^{-4}$.
 }
 \label{fig:noise}
\end{figure}

{Estimation of the internal transition probability of the frequency converter}.
We estimated the internal transition probability of the frequency converter
by constructing a theoretical model as follows. 
We assume that 
the optical circuit of the frequency-domain HOM interferometer
is decomposed to a lossless frequency converter, two loss media inducing the loss in each of the upper and the lower input modes, and two spectral filters acting on each of two output modes.
The transition probability and the staying probability of the lossless frequency converter
are described by $R=R(P,\omega)$ and $T = 1-R(P,\omega)$.
The transmittances of the loss media and the spectral filters for the upper/lower light
are denoted by $T_{\mathrm{in, U/L}}$ and $T_{\mathrm{out, U/L}}(\omega)$, respectively. 
As is described in the main text,
we assume the state of the input light composed of the single photon and the coherent state
is $\ket{\phi}=\hat{a}^\dagger_{\mathrm{U}}\hat{D}_{\mathrm{L}}(\alpha)\ket{0}$, 
where $\hat{a}^\dagger_{\mathrm{U}}$ is a creation operator of the upper mode,
 $\hat{D}_{\mathrm{L}}(\alpha)$ is a displacement operator with complex number $\alpha$ of the lower mode
  and $\ket{0}$ is the vacuum state for both modes.
The spectral shapes of the heralded single photon and the coherent light pulse
 are denoted by $F_{\mathrm{U}}(\omega)$ and $F_{\mathrm{L}}(\omega)$, respectively.
They are normalized as $\int d\omega F_{\mathrm{U}}(\omega)=\int d\omega F_{\mathrm{L}}(\omega)=1$.

From the theoretical model, the observed values in Fig.~2 are described by
\begin{eqnarray}
p_{\mathrm{U,t}}(P) &=& T_{\mathrm{in, U}} \int d \omega F_{\mathrm{U}}(\omega) R(P, \omega) T_{\mathrm{out, L}}(\omega),\\
p_{\mathrm{U,s}}(P) &=& T_{\mathrm{in, U}} \int d \omega F_{\mathrm{U}}(\omega) (1-R(P, \omega)) T_{\mathrm{out, U}}(\omega),\\
p_{\mathrm{L,t}}(P) &=& |\alpha|^2 T_{\mathrm{in, L}} \int d \omega F_{\mathrm{L}}(\omega) R(P, \omega) T_{\mathrm{out, U}}(\omega),\\
p_{\mathrm{L,s}}(P) &=& |\alpha|^2 T_{\mathrm{in, L}} \int d \omega F_{\mathrm{L}}(\omega) (1-R(P, \omega)) T_{\mathrm{out, L}}(\omega).
\end{eqnarray}
We assume that $F_{\mathrm{U}}(\omega)$, $F_{\mathrm{L}}(\omega)$ and $R(P,\omega)$
 are Gaussian with bandwidths of $\Delta_{\mathrm{in,U}}= 740$~GHz, $\Delta_{\mathrm{in,L}}=93$~GHz and $\Delta_{\mathrm{WG}}=140$~GHz, respectively.
The peak value of $R(P,\omega)$ is described by $\widetilde{R}(P)$.
We also assume that $T_{\mathrm{out, U}}(\omega)$ and $T_{\mathrm{out, L}}(\omega)$ are
Gaussian with the bandwidths of $\Delta_{\mathrm{out,U}}=70$~GHz and $\Delta_{\mathrm{out,L}}=92$~GHz around
the center of 780~nm/1522~nm for the upper/lower input light,
at which the peak values are given by $\widetilde{T}_{\mathrm{out, U}}$ and $\widetilde{T}_{\mathrm{out, L}}$.
From the four observed values $p_{\mathrm{U,t}}(P), p_{\mathrm{U,s}}(P),$ $p_{\mathrm{L,t}}(P)$ and $p_{\mathrm{L,s}}(P)$
in Fig.~2,
we obtain the four parameters $\widetilde{R}(P)$,  $T_{\mathrm{in, U}}\widetilde{T}_{\mathrm{out, U}}$,
$T_{\mathrm{in, U}}\widetilde{T}_{\mathrm{out, L}}$ and $|\alpha|^2 T_{\mathrm{in, L}}/T_{\mathrm{in, U}}$
for various values of $P$. 
The four observed values are described by
\begin{eqnarray}
p_{\mathrm{U,t}}(P) 
&=& T_{\mathrm{in, U}}\widetilde{R}(P) \widetilde{T}_{\mathrm{out,L}}
\frac{1}
{\sqrt{1+\Delta_{\mathrm{in,U}}^2/\Delta_{\mathrm{WG}}^2
+\Delta_{\mathrm{in,U}}^2/\Delta_{\mathrm{out,L}}^2}},\label{eq:put}\\
p_{\mathrm{U,s}}(P) 
&=& T_{\mathrm{in, U}} \widetilde{T}_{\mathrm{out,U}}\left(
\frac{1}
{\sqrt{1+\Delta_{\mathrm{in,U}}^2/\Delta_{\mathrm{out,U}}^2}}
- \widetilde{R}(P)
\frac{1}
{\sqrt{1+\Delta_{\mathrm{in,U}}^2/\Delta_{\mathrm{WG}}^2
+\Delta_{\mathrm{in,U}}^2/\Delta_{\mathrm{out,U}}^2}}\right),\label{eq:pus}\\
p_{\mathrm{L,t}}(P) 
&=& |\alpha|^2T_{\mathrm{in, L}} \widetilde{R}(P) \widetilde{T}_{\mathrm{out,U}}
\frac{1}
{\sqrt{1+\Delta_{\mathrm{in,L}}^2/\Delta_{\mathrm{WG}}^2
+\Delta_{\mathrm{in,L}}^2/\Delta_{\mathrm{out,U}}^2}}\label{eq:plt},\\
p_{\mathrm{L,s}}(P) 
&=&|\alpha|^2T_{\mathrm{in, L}}  \widetilde{T}_{\mathrm{out,L}}\left(
\frac{1}
{1+\sqrt{\Delta_{\mathrm{in,L}}^2/\Delta_{\mathrm{out,L}}^2}}
- \widetilde{R}(P) 
\frac{1}
{\sqrt{1+\Delta_{\mathrm{in,L}}^2/\Delta_{\mathrm{WG}}^2
+\Delta_{\mathrm{in,L}}^2/\Delta_{\mathrm{out,L}}^2}}\right).\label{eq:pls}
\end{eqnarray}
From the above equations, we obtain the quadratic equation 
$0= A \widetilde{R}(P)^2 +B\widetilde{R}(P) +C$, where
\begin{eqnarray}
A &\equiv&
\frac{1}
  {
    \sqrt{(1+\Delta_{\mathrm{in,U}}^2/\Delta_{\mathrm{WG}}^2
+\Delta_{\mathrm{in,U}}^2/\Delta_{\mathrm{out,L}}^2) (
   1+\Delta_{\mathrm{in,L}}^2/\Delta_{\mathrm{WG}}^2
+\Delta_{\mathrm{in,L}}^2/\Delta_{\mathrm{out,U}}^2 )}
  }\nonumber\\
  &&-\frac{1}
  {
    \sqrt{(1+\Delta_{\mathrm{in,U}}^2/\Delta_{\mathrm{WG}}^2
+\Delta_{\mathrm{in,U}}^2/\Delta_{\mathrm{out,U}}^2) (
   1+\Delta_{\mathrm{in,L}}^2/\Delta_{\mathrm{WG}}^2
+\Delta_{\mathrm{in,L}}^2/\Delta_{\mathrm{out,L}}^2 )}
  }\frac{p_{\mathrm{U,t}}(P) p_{\mathrm{L,s}}(P)}{p_{\mathrm{U,s}}(P) p_{\mathrm{L,t}}(P)},\\
B &\equiv&
\left(
 \frac{1}
{\sqrt{(1+\Delta_{\mathrm{in,U}}^2/\Delta_{\mathrm{out,U}}^2)
(1+\Delta_{\mathrm{in,L}}^2/\Delta_{\mathrm{WG}}^2
+\Delta_{\mathrm{in,L}}^2/\Delta_{\mathrm{out,L}}^2)}}
\right.\nonumber\\
&&\left.+
\frac{1}
{\sqrt{(1+\Delta_{\mathrm{in,L}}^2/\Delta_{\mathrm{out,L}}^2)
(1
+\Delta_{\mathrm{out,L}}^2/\Delta_{\mathrm{WG}}^2
+\Delta_{\mathrm{out,L}}^2/\Delta_{\mathrm{in,L}}^2)
}}\right)
 \frac{p_{\mathrm{U,t}}(P) p_{\mathrm{L,s}}(P)}{p_{\mathrm{U,s}}(P) p_{\mathrm{L,t}}(P)},\\
C&\equiv&
-\frac{1}
{\sqrt{(1+\Delta_{\mathrm{in,U}}^2/\Delta_{\mathrm{out,U}}^2)
(1+\Delta_{\mathrm{in,L}}^2/\Delta_{\mathrm{out,L}}^2)}}
 \frac{p_{\mathrm{U,t}}(P) p_{\mathrm{L,s}}(P)}{p_{\mathrm{U,s}}(P) p_{\mathrm{L,t}}(P)}.
\end{eqnarray}
We see that 
$\widetilde{R}(P)=\frac{-B+\sqrt{B^2-4AC}}{2A}$ is the physical solution which satisfies $0\leq \widetilde{R}(P) \leq 1$.
The result of $\widetilde{R}(P)$ is shown in Fig.~3.
The other three parameters $T_{\mathrm{in, U}}\widetilde{T}_{\mathrm{out, U}}$,
$T_{\mathrm{in, U}}\widetilde{T}_{\mathrm{out, L}}$ and $|\alpha|^2 T_{\mathrm{in, L}}/T_{\mathrm{in, U}}$
are described by
\begin{eqnarray}
\frac{|\alpha|^2T_{\mathrm{in, L}} }{T_{\mathrm{in, U}}}
&=&
\frac{p_{\mathrm{L,s}}(P)}{p_{\mathrm{U,t}}(P)}\frac{\widetilde{R}(P)}{\sqrt{1
+\Delta_{\mathrm{in,U}}^2/\Delta_{\mathrm{WG}}^2
+\Delta_{\mathrm{in,U}}^2/\Delta_{\mathrm{out,L}}^2
}}\nonumber\\
&&\times\left(\frac{1}
{\sqrt{1+\Delta_{\mathrm{in,L}}^2/\Delta_{\mathrm{out,L}}^2}}
- \widetilde{R}(P)
\frac{1}
{\sqrt{1
+\Delta_{\mathrm{in,L}}^2/\Delta_{\mathrm{WG}}^2
+\Delta_{\mathrm{in,L}}^2/\Delta_{\mathrm{out,L}}^2
}}\right)^{-1},\\
T_{\mathrm{in,U}}\widetilde{T}_{\mathrm{out,U}}
&=& p_{\mathrm{U,s}}(P) \left(
\frac{1}
{\sqrt{1+\Delta_{\mathrm{in,U}}^2/\Delta_{\mathrm{out,U}}^2}}
- \widetilde{R}(P)
\frac{1}
{\sqrt{1+\Delta_{\mathrm{in,U}}^2/\Delta_{\mathrm{WG}}^2
+\Delta_{\mathrm{in,U}}^2/\Delta_{\mathrm{out,U}}^2}}\right)^{-1},\\
T_{\mathrm{in,U}}\widetilde{T}_{\mathrm{out,L}}
&=&  p_{\mathrm{U,t}}(P)
\frac{\sqrt{1
+\Delta_{\mathrm{in,U}}^2/\Delta_{\mathrm{WG}}^2
+\Delta_{\mathrm{in,U}}^2/\Delta_{\mathrm{out,L}}^2
}}{\widetilde{R}(P)}.
\end{eqnarray} 
From these equations, we calculated these parameters by using the observed values.
The result is shown in Fig.~\ref{fig:precalc}.
We note that these parameters
should be independent of the pump power.
The estimated values of $T_{\mathrm{in, U}}\widetilde{T}_{\mathrm{out, U}}$,
 $T_{\mathrm{in, U}}\widetilde{T}_{\mathrm{out, L}}$ and $|\alpha|^2 T_{\mathrm{in, L}}/T_{\mathrm{in, U}}$ take about constant values of 0.078, 0.081 and 0.047, respectively,  which is consistent with our theoretical model.
By using these estimated values, we obtain the dashed curves in Fig.~2.
These curves are in good agreement with the experimental results.

 {Estimation of the degradation of the visibility}.
We estimated the degradation of the visibility
by using the theoretical model as follows. 
The coincidence probability $p_c(\tau)$ on the time delay $\tau$ is described by
\begin{eqnarray}
p_c(\tau)
  &=&1-p_\mathrm{U0}(\tau)- p_\mathrm{L0}(\tau)+ p_\mathrm{U0,L0}(\tau), \label{eq:viscal}
\end{eqnarray}
where 
$p_\mathrm{U0/L0}(\tau)$ and $p_\mathrm{U0,L0}(\tau)$ represent 
probabilities where no photon is detected on the time delay $\tau$
 in the upper/lower mode and in both modes, respectively. 
The visibility of the HOM interference is defined by $1-p_c(0)/p_c(\infty)$.
In our theoretical model, 
these probabilities are described by
\begin{eqnarray}
p_\mathrm{U0}(\tau)&=&(1-d_{\mathrm{U}})\Vert \bra{0}_{\mathrm{U}}  \hat{U}_\mathrm{FBS}\ket{\phi}_\mathrm{U,L}\otimes\ket{0}_{\mathrm{E}}\Vert^2,\label{eq:pu0}\\
p_\mathrm{L0}(\tau)&=&(1-d_{\mathrm{L}})\Vert \bra{0}_{\mathrm{L}}  \hat{U}_\mathrm{FBS}\ket{\phi}_\mathrm{U,L}\otimes\ket{0}_{\mathrm{E}}\Vert^2,\label{eq:pl0}\\
p_\mathrm{U0,L0}(\tau)&=&(1-d_{\mathrm{U}})(1-d_{\mathrm{L}})\Vert \bra{0}_{\mathrm{U,L}}  \hat{U}_\mathrm{FBS}\ket{\phi}_\mathrm{U,L}\otimes\ket{0}_{\mathrm{E}}\Vert^2,\label{eq:pul0}
\end{eqnarray}
where
$d_{\mathrm{U/L}}$ is the observed detection probability of the background noise, shown in Fig.~\ref{fig:noise}.
$\ket{\phi}$ is the initial state described by $\hat{a}^\dagger_{\mathrm{U}}\hat{D}_{\mathrm{L}}(\alpha)\ket{0}$.
$\ket{0}_{\mathrm{E}}$ is the vacuum state in an ancillary system describing the input/output loss modes.
$\hat{U}_\mathrm{FBS}$ is a unitary operator describing the frequency-domain BS including the input and output loss.
We decompose $\hat{U}_\mathrm{FBS}$ to five lossless components, one of which describes 
lossless part of the frequency-domain BS and others describe 
the lossy media/spectral filters acting on input/output modes.
The actions of $\hat{U}_\mathrm{FBS}$ on the creation operators  
$\hat{a}^\dagger_{\mathrm{U}}=\int d\omega \sqrt{F_{\mathrm{U}}(\omega)}\hat{a}^\dagger_{\mathrm{U},\omega}$ 
in the input upper mode and 
$\hat{a}^\dagger_{\mathrm{L}}=\int d\omega e^{-i \omega \tau}\sqrt{F_{\mathrm{L}}(\omega)}\hat{a}^\dagger_{\mathrm{L},\omega}$ in the input lower mode are written by 
\begin{eqnarray}
  \hat{U}_\mathrm{FBS} \hat{a}_\mathrm{U}^\dagger\hat{U}_\mathrm{FBS}^\dagger
    &=& 
    \sqrt{R_{\mathrm{in, U}}} \int d \omega \sqrt{F_{\mathrm{U}}(\omega)} \hat{a}^\dagger_{\mathrm{EU1},\omega}\nonumber\\
&&+\sqrt{T_{\mathrm{in, U}}} \int d \omega \sqrt{F_{\mathrm{U}}(\omega) T(P, \omega)}\left(\sqrt{T_{\mathrm{out, U}}(\omega)}\hat{a}^\dagger_{\mathrm{U},\omega}+
\sqrt{R_{\mathrm{out, U}}(\omega)}\hat{a}^\dagger_{\mathrm{EU2},\omega}\right)\nonumber\\
&&+\sqrt{T_{\mathrm{in, U}}} \int d \omega e^{-i\phi}\sqrt{F_{\mathrm{U}}(\omega) R(P, \omega)}
  \left(\sqrt{ T_{\mathrm{out, L}}(\omega)}\hat{a}^\dagger_{\mathrm{L},\omega}+\sqrt{ R_{\mathrm{out, L}}(\omega)}\hat{a}^\dagger_{\mathrm{EL2},\omega}\right),\\
  \hat{U}_\mathrm{FBS} \hat{a}_\mathrm{L}^\dagger\hat{U}_\mathrm{FBS}^\dagger
    &=& 
    \sqrt{R_{\mathrm{in, L}}} \int d \omega e^{-i \omega \tau}\sqrt{F_{\mathrm{L}}(\omega)} \hat{a}^\dagger_{\mathrm{EL1},\omega}\nonumber\\
&&+\sqrt{T_{\mathrm{in, L}}} \int d \omega e^{-i \omega \tau}\sqrt{F_{\mathrm{L}}(\omega) T(P, \omega) } \left(\sqrt{T_{\mathrm{out, L}}(\omega)}\hat{a}^\dagger_{\mathrm{L},\omega}+\sqrt{R_{\mathrm{out, L}}(\omega)}\hat{a}^\dagger_{\mathrm{EL2},\omega}\right)\nonumber\\
&&-\sqrt{T_{\mathrm{in, L}}} \int d \omega e^{-i \omega \tau+i\phi}\sqrt{F_{\mathrm{L}}(\omega) R(P, \omega)} \left(\sqrt{ T_{\mathrm{out, U}}(\omega)}\hat{a}^\dagger_{\mathrm{U},\omega}+\sqrt{ R_{\mathrm{out, U}}(\omega)}\hat{a}^\dagger_{\mathrm{EU2},\omega}\right),
\end{eqnarray} 
where $R_\mathrm{in,U/L}=1-T_\mathrm{in,U/L}, R_{\mathrm{out, U/L}}(\omega)=1-T_{\mathrm{out, U/L}}(\omega)$ and 
$[\hat{a}_{i,\omega}, \hat{a}_{j,\omega'}^\dagger] =\delta_{i,j}\delta(\omega-\omega') $ for
$i,j = \mathrm{U,L,EU1,EU2,EL1,EL2}$ .
From Eqs.~(\ref{eq:pu0}) -- (\ref{eq:pul0}), we have
\begin{eqnarray}
p_\mathrm{U0,L0}(\tau)&=& (1-d_{\mathrm{U}})(1-d_{\mathrm{L}})
	\left\Vert \sqrt{R_{\mathrm{in, U}}} \int d \omega \sqrt{F_{\mathrm{U}}(\omega)}      
        \hat{a}^\dagger_{\mathrm{EU1},\omega}  \bra{0}_\mathrm{U,L}\hat{U}_\mathrm{FBS} 
        \hat{D}_{\mathrm{L}}(\alpha)\ket{0}_\mathrm{U,L,E}\right\Vert^2 \nonumber\\
     &&+
        (1-d_{\mathrm{U}})(1-d_{\mathrm{L}})\nonumber\\
     &&\left\Vert \sqrt{T_{\mathrm{in, U}}} \int d \omega \sqrt{F_{\mathrm{U}}(\omega) R(P, \omega)}
        (\sqrt{R_{\mathrm{out, U}}(\omega)}     
        \hat{a}^\dagger_{\mathrm{EU2},\omega}
        +e^{-i\phi}\sqrt{R_{\mathrm{out, L}}(\omega)}\hat{a}^\dagger_{\mathrm{EL2},\omega}) \right. \nonumber\\
     &&\times   \left. \bra{0}_\mathrm{U,L}\hat{U}_\mathrm{FBS} 
        \hat{D}_{\mathrm{L}}(\alpha)\ket{0}_\mathrm{U,L,E}\right\Vert^2 \\
&=& (1-d_{\mathrm{U}})(1-d_{\mathrm{L}})\nonumber\\
    &&\times\exp\left(
        - |\alpha|^2\frac{T_{\mathrm{in, L}}}{T_{\mathrm{in, U}}} \int d \omega F_{\mathrm{L}}(\omega) (R(P, \omega)  
        T_{\mathrm{in, U}}T_{\mathrm{out, U}}(\omega)+T(P, \omega) T_{\mathrm{in, U}}
        T_{\mathrm{out, L}}(\omega) ) \right)\nonumber\\
    &&\times\Biggl(
        \int d \omega F_{\mathrm{U}}(\omega) \left(1- T(P, \omega)T_{\mathrm{in, U}} 
        T_{\mathrm{out, U}}(\omega)-R(P, \omega)T_{\mathrm{in, U}} T_{\mathrm{out, L}}(\omega)\right)\nonumber\\
    &&+
        |\alpha|^2\frac{T_{\mathrm{in, L}}}{T_{\mathrm{in, U}}}
         \left|\int d \omega e^{-i\omega \tau}\sqrt{F_{\mathrm{U}}(\omega)F_{\mathrm{L}}(\omega) 
         T(P, \omega)R(P, \omega)}T_{\mathrm{in, U}}
         (T_{\mathrm{out, U}}(\omega)-T_{\mathrm{out, L}}(\omega))\right|^2\Biggr),\label{eq:pul}
         \end{eqnarray}
         \begin{eqnarray}
p_\mathrm{U0}(\tau)&=& (1-d_{\mathrm{U}})\exp\left(
         - |\alpha|^2\frac{T_{\mathrm{in, L}}}{T_{\mathrm{in, U}}} \int d \omega F_{\mathrm{L}}(\omega) 
         R(P, \omega)T_{\mathrm{in, U}}T_{\mathrm{out, U}}(\omega)\right)\nonumber\\
&&   \times\Biggl(
         \int d \omega F_{\mathrm{U}}(\omega) (1-T(P, \omega) T_{\mathrm{in, U}} T_{\mathrm{out, U}}(\omega))\nonumber\\
    &&+
        |\alpha|^2\frac{T_{\mathrm{in, L}}}{T_{\mathrm{in, U}}}
        \left|\int d \omega e^{-i\omega \tau}\sqrt{F_{\mathrm{U}}(\omega)F_{\mathrm{L}}(\omega) T(P, \omega)R(P, \omega)}
        T_{\mathrm{in, U}}T_{\mathrm{out, U}}(\omega)\right|^2\Bigg),\label{eq:pu}\\
p_\mathrm{L0}(\tau)&=&(1-d_{\mathrm{L}}) \exp\left(
        - |\alpha|^2\frac{T_{\mathrm{in, L}}}{T_{\mathrm{in, U}}}
         \int d \omega F_{\mathrm{L}}(\omega) T(P, \omega) T_{\mathrm{in, U}}T_{\mathrm{out, L}}(\omega)\right)\nonumber\\
    &&\times\Biggl(
        \int d \omega F_{\mathrm{U}}(\omega) (1-R(P, \omega) T_{\mathrm{in, U}} T_{\mathrm{out, L}}(\omega))\nonumber\\
    &&+
        |\alpha|^2 \frac{T_{\mathrm{in, L}}}{T_{\mathrm{in, U}}}
        \left|\int d \omega e^{-i\omega \tau}\sqrt{F_{\mathrm{U}}(\omega)F_{\mathrm{L}}(\omega) T(P, \omega)R(P, \omega)}T_{\mathrm{in, U}}T_{\mathrm{out, L}}(\omega)\right|^2\Bigg).\label{eq:pl}
\end{eqnarray}
By using the experimental parameters $\widetilde{R}(P)$,  $T_{\mathrm{in, U}}\widetilde{T}_{\mathrm{out, U}}$,
 $T_{\mathrm{in, U}}\widetilde{T}_{\mathrm{out, L}}$ and $|\alpha|^2 T_{\mathrm{in, L}}/T_{\mathrm{in, U}}$,
we obtain the dashed curves in Figs.~4a and 4b from Eq.~(\ref{eq:viscal}).
Fig.~4b indicates that the highest visibility of 0.81 will be obtained at 190-mW pump power.

In our theoretical model,
main reason for the degradation of the visibility comes from input light pulses;
the effect of the multiphoton components in the coherent light pulse at 1522~nm
and the broad bandwidth of $F_\mathrm{U}(\omega)$.
When we replace the coherent light pulse by a single photon 
 the visibility is calculated in a similar way with $\ket{\phi}=\hat{a}^\dagger_{\mathrm{U}}\hat{a}^\dagger_{\mathrm{L}} \ket{0}_\mathrm{U,L}$.
In this case, the probabilities where no photon is detected in both modes $p^\mathrm{ph}_\mathrm{U0,L0}(\tau)$ and 
 in the upper/lower mode $p^\mathrm{ph}_\mathrm{U0/L0}(\tau)$
are described by
\begin{eqnarray}
p^\mathrm{ph}_\mathrm{U0,L0}(\tau)&\equiv&(1-d_{\mathrm{U}})(1-d_{\mathrm{L}})\Vert \bra{0}_{\mathrm{U,L}}  
        \hat{U}_\mathrm{FBS}\hat{a}^\dagger_{\mathrm{U}}\hat{a}^\dagger_{\mathrm{L}} \ket{0}_\mathrm{U,L,E}\Vert^2\nonumber\\
    &=&(1-d_{\mathrm{U}})(1-d_{\mathrm{L}})
        \int d \omega F_{\mathrm{U}}(\omega) \left(1- T(P, \omega)T_{\mathrm{in, U}} 
        T_{\mathrm{out, U}}(\omega)-R(P, \omega)T_{\mathrm{in, U}} T_{\mathrm{out, L}}(\omega)\right)\nonumber\\
    &&\times
        \frac{T_{\mathrm{in, L}}}{T_{\mathrm{in, U}}}
        \int d \omega F_{\mathrm{L}}(\omega) \left(1- R(P, \omega)T_{\mathrm{in, U}} 
        T_{\mathrm{out, U}}(\omega)-T(P, \omega)T_{\mathrm{in, U}} T_{\mathrm{out, L}}(\omega)\right)\nonumber\\
    &&+(1-d_{\mathrm{U}})(1-d_{\mathrm{L}})\nonumber\\
    &&\times
         \frac{T_{\mathrm{in, L}}}{T_{\mathrm{in, U}}}
         \left|\int d \omega e^{-i\omega \tau}\sqrt{F_{\mathrm{U}}(\omega)F_{\mathrm{L}}(\omega) 
         T(P, \omega)R(P, \omega)}T_{\mathrm{in, U}}
         (T_{\mathrm{out, U}}(\omega) -T_{\mathrm{out, L}}(\omega))\right|^2,\label{eq:phpul}\\
p^\mathrm{ph}_\mathrm{U0}(\tau)&\equiv&(1-d_{\mathrm{U}})\Vert \bra{0}_{\mathrm{U}}  \hat{U}_\mathrm{FBS}\hat{a}^\dagger_{\mathrm{U}}\hat{a}^\dagger_{\mathrm{L}} \ket{0}_\mathrm{U,L,E}\Vert^2\nonumber\\
    &=&(1-d_{\mathrm{U}})
        \int d \omega F_{\mathrm{U}}(\omega) \left(1- T(P, \omega)T_{\mathrm{in, U}} 
        T_{\mathrm{out, U}}(\omega)\right)\nonumber\\
    &&\times
        \frac{T_{\mathrm{in, L}}}{T_{\mathrm{in, U}}}
        \int d \omega F_{\mathrm{L}}(\omega) \left(1- R(P, \omega)T_{\mathrm{in, U}} 
        T_{\mathrm{out, U}}(\omega)\right)\nonumber\\
    &&+(1-d_{\mathrm{U}})
         \frac{T_{\mathrm{in, L}}}{T_{\mathrm{in, U}}}
         \left|\int d \omega e^{-i\omega \tau}\sqrt{F_{\mathrm{U}}(\omega)F_{\mathrm{L}}(\omega) 
         T(P, \omega)R(P, \omega)}T_{\mathrm{in, U}}T_{\mathrm{out, U}}(\omega)\right|^2,\label{eq:phpu}\\
p^\mathrm{ph}_\mathrm{L0}(\tau)&\equiv&(1-d_{\mathrm{L}})\Vert \bra{0}_{\mathrm{U,L}}  \hat{U}_\mathrm{FBS}\hat{a}^\dagger_{\mathrm{U}}\hat{a}^\dagger_{\mathrm{L}} \ket{0}_\mathrm{U,L,E}\Vert^2 \nonumber\\
    &=&(1-d_{\mathrm{L}})
        \int d \omega F_{\mathrm{U}}(\omega) \left(1- R(P, \omega)T_{\mathrm{in, U}} 
        T_{\mathrm{out, L}}(\omega)\right)\nonumber\\
    &&\times
        \frac{T_{\mathrm{in, L}}}{T_{\mathrm{in, U}}}
        \int d \omega F_{\mathrm{L}}(\omega) \left(1- T(P, \omega)T_{\mathrm{in, U}} 
        T_{\mathrm{out, L}}(\omega)\right)\nonumber\\
    &&+(1-d_{\mathrm{L}})
         \frac{T_{\mathrm{in, L}}}{T_{\mathrm{in, U}}}
         \left|\int d \omega e^{-i\omega \tau}\sqrt{F_{\mathrm{U}}(\omega)F_{\mathrm{L}}(\omega) 
         T(P, \omega)R(P, \omega)}T_{\mathrm{in, U}}T_{\mathrm{out, L}}(\omega)\right|^2.\label{eq:phpl}
\end{eqnarray}
When we adjust $T_\mathrm{in,L}$ such that  $T_\mathrm{in,L}/T_\mathrm{in,U}=1$ is satisfied,
by using the experimental parameters $\widetilde{R}(P)$,  $T_{\mathrm{in, U}}\widetilde{T}_{\mathrm{out, U}}$ and
$T_{\mathrm{in, U}}\widetilde{T}_{\mathrm{out, L}}$, 
we obtain the visibility of 0.95 at 190-mW pump power from Eqs.~(\ref{eq:phpul}) -- (\ref{eq:phpl}).
In addition, if we narrow the bandwidth of $F_\mathrm{U}(\omega)$ from $\Delta_{\mathrm{in,U}}$ to $\Delta_{\mathrm{in,L}}$, 
the visibility will be 0.98 at 190-mW pump power from Eqs.~(\ref{eq:phpul}) -- (\ref{eq:phpl}).
We note that if we take only the improvement of the bandwidth,
the visibility will be 0.93 at 190-mW pump power from Eqs.~(\ref{eq:pul}) -- (\ref{eq:pl}).

\end{document}